# EQUIVALENT SOURCES APPROACH FOR NONRADIATING STATES AND ANAPOLE METAMATERIALS

Giuseppe Labate[1], Anar K. Ospanova[2], Nikita A. Nemkov[2,3], Alexey A. Basharin[2,4], Ladislau Matekovits[1,5]

[1] Politecnico di Torino, Department of Electronic and Telecommunications, 10129, Torino, Italy
[2] National University of Science and Technology (MISiS), The Laboratory of Superconducting metamaterials, 119049, Moscow, Russia
[3] Moscow Institute of Physics and Technology (MIPT), 141701, Moscow region, Russia
[4] National University of Science and Technology (MISiS), Department of Theoretical Physics and Quantum Technologies, 119049, Moscow, Russia
[5] Faculty of Science, Macquarie University, Sydney, NSW 2109, Australia

Correspondence and requests should be addressed to Anar K. Ospanova (anar.k.ospanova@gmail.com)

**Abstract**: In this work, we discuss theoretical findings on the common feature describing nonradiating sources, based on equivalent sources from which it is possible to derive cloaking devices and anapole mode conditions. Starting from the differential form of Maxwell's Equations expressed in terms of equivalent electromagnetic sources, we derive two unique compact conditions. By specifying the nature of these passively induced or actively impressed current density sources in certain volumes and/or predefined surfaces, we derive theoretical results consistently with the literature about nonradiating particles, cloaking devices and anapole mode structures through peculiar destructive interactions between volumetric-volumetric, volumetric-surface and surface-surface equivalent sources.

## I. NONRADIATING SOURCES AND CLOAKING DEVICES

Nonradiating artificial configurations attract more attention as candidates for invisible models with analogy with stable atoms [1]. One of the possible solutions is meta-particles with anapole modes excitation. In particular, so-called toroidal dipole and electric dipole families have the same angular distribution and parity properties. Correspondingly, the electromagnetic fields of destructively oscillated electric and toroidal dipoles disappear everywhere apart from the origin, formed anapole nonradiating mode [2]. The interest in anapole nonradiating sources arise in the context of inverse scattering problem of electrodynamics, in order to reconstruction of source from radiated fields, suppression of scattering and invisibility problems in nanophotonics and metamaterials [3]. On can define nonradiating source as an oscillating current configuration of a finite size, which does not generate any fields outside of its occupied volume.

Eliminating the electromagnetic scattering and radiation outside a certain volume of interest *V*, bounded by its surface *Γ*, is the common feature of all nonradiating sources by definition, both in the single-core as in the core-shell architectures [4-7]. Since the early work of Devaney and Wolf on five nonradiating theorems [8] the stimulated contributions on setting the problem for nonradiating currents have been employed for ensuring these trapped modes within engineered particles, metamaterial structures and metasurface devices for several applications as for example the reduction of mutual interferences between nearby antennas [9, 10] and nonradiating anapole states [11-15]. However, a proper understanding of these nonradiating modes turn out to be useful also for related radiative energy features expressed in terms of transverse electric (TE) and transverse magnetic (TM) radiating modes [16] or in terms of proper degrees of freedom for imposing specific phase values at the surface of a reflectarray antenna [17]. The possibility to discriminate in the volume $V$ between a subdomain $V_1$, where a fixed object is placed, and a subdomain $V_2$, where an additional tunable controlling layer is inserted, has given the possibility to enforce a nonradiating condition in $\Gamma$ even when the object by itself scatters energy outside $V_1$ due to a destructive interference with the $V_2$ region: this has given rise to several *cloaking* techniques and devices [18-20].

Among all the cloaking techniques that exploit, for the suppression of electromagnetic scattering, plasmonic materials [6], anisotropic metamaterials [18] or metasurface coatings [7], the synthesis of a surface impedance function in the proximity of an object has been recently highlighted in order to sustain a nonradiating feature with a straightforward design for waveguides [21], metallic cylinders [22] and a general scattering phenomenon [20]. As a counter-example, Transformation Optics features in rerouting the incident electromagnetic wave around the object, eliminating the direct interaction between them, i.e., the scattering [24, 25].

Among all the nonradiating configurations, recent interests in the research community have arisen for the explanation of the phenomenon in terms of *anapole mode*. This concentrated near field distribution has been experimentally observed in dielectric nanodisks [11] and in 3D SRR-like structures [14]. The results of this subject [12, 13, 26-28] are underlying the nowadays-strong interest towards the nonradiating theory.

Inspired by the equivalence theorem approach for the surface impedance cloaking technique [23], we show how the destructive interference between surface equivalent sources can be extended to the anapole mode, as first proposed in relation with only Transformation Optics approaches [24]. Here we assume the destructive interference between surface equivalent sources explained in terms of the toroidal and electric dipole moments.

Starting from the differential form of the Maxwell's Equations, we derive two compact conditions, that accurately described the nonradiating phenomenon. Moreover, we emphasize the nonradiating particles in terms of volumetric magnetization and polarization vectors, resulting in anapole modes. Realistic structures exhibiting transparrensy effect are devised and numerically simulated, demonstrating the validity of the proposed paradigm.

## II. HIDDEN NONRADIATING EQUIVALENT SOURCES IN DIFFERENTIAL MAXWELL'S EQUATIONS

Let us discuss the graphical representation of any scatterer separated on the overall volume $V$ in two subdomains: a generic cluster of volumetric scatterers – represented by $V_1$, and the additional subdomain $V_2$ (Fig. 1). The surface currents excited by the incident electromagnetic wave generate surface and volumetric currents that in turn will radiate. The main feature of such nonradiating configuration is related to the destructive interference between the currents $J_1$ and $J_2$ on/in the object to be hidden, i.e., in $V_1$, and those on/in the properly positioned additional matter, i.e., in $V_2$. In case the internal element of volume $V_1$ supports (an electric), dipole-like response described in terms of its polarization vector **P**, the additional structure of volume $V_2$ should produce a current distribution that will generate a similar radiation pattern. In case this is done by another electric dipole-like structure, their far-field radiations should possess destructive interference to each other, i.e., giving a rise to the scattering cancellation configuration. However, other current arrangements radiate in a similar manner. As for example a toroidal one that is characterized by lateral poloidal currents sustaining a toroidization vector *T* [3, 14, 29]. If properly excited, the two radiations (*P* and *T*) will cancel out each other, bringing out a non-radiating configuration in the far field.

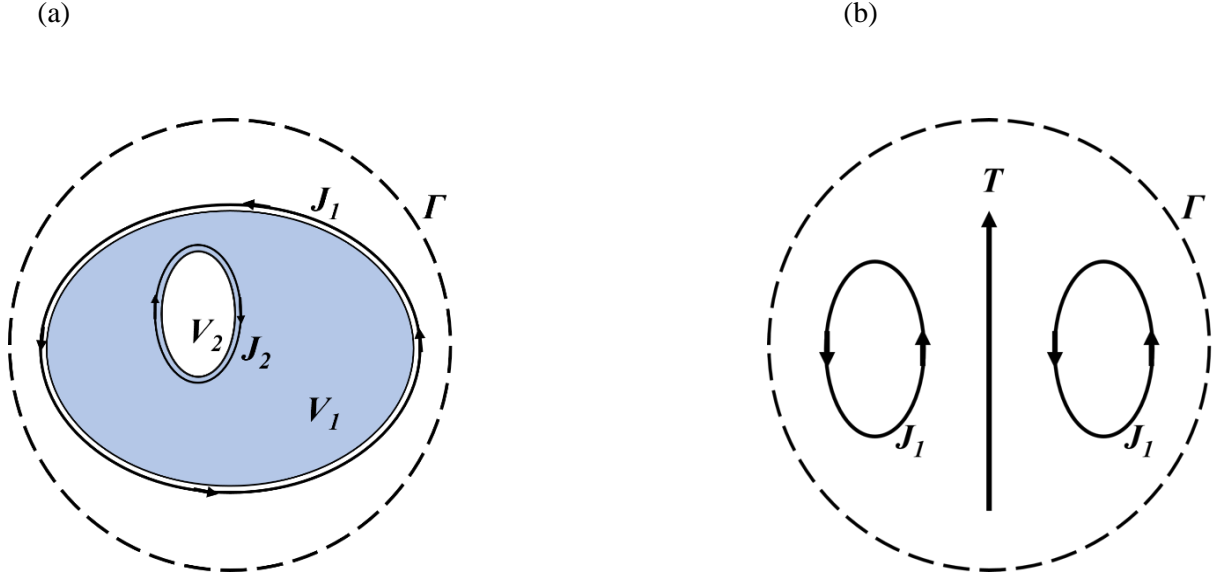

Fig. 1. Volume $V_1$ incorporating a generic cluster $V_2$ of volumetric dielectric scatterer (a). $\Gamma$ indicates the virtual boundary (surface) between the V and the external space. Equivalent treatment (b) with volumetric dielectric sources $J_1$ and $J_2$.

We start with the differential Maxwell's Equations, with right hand term as equivalent sources (that in the most general case includes both surface and volumetric contributions), expressed as follows

$$\nabla \times \mathbf{E} = -i\omega\mu_0 \mathbf{H} \tag{1}$$

$$\nabla \times \mathbf{H} = +i\omega\varepsilon_0 \mathbf{E} + \mathbf{J}_{eq} \tag{2}$$

where $\omega$ denotes the angular frequency, and the $e^{-i\omega t}$ time convention has been assumed. Consecutively, the equivalent (electromagnetic) source $\mathbf{J}_{eq}$ is expressed as a combination of the electric and/or magnetic volumetric and surface current components. In particular, one can express [31] these terms for the *volumetric* contributions, as follows:

$$- \quad J_v^e(r') = +i\omega\varepsilon_0 \mathbf{P}(r') \qquad \text{with} \quad r' \in V \tag{3}$$
$$- \quad J_v^m(r') = -Y_0 \nabla \times \mathbf{M}(r') \qquad \text{with} \quad r' \in V \tag{4}$$

where **P** and **M** are the polarization and magnetization vectors, respectively. $Y_0 = \sqrt{\varepsilon_0/\mu_0}$ indicates the intrinsic vacuum admittance (here considering vacuum as reference background).

As concern the equivalent *surface* currents, they have the expressions

$$J_s^e(r') = +\hat{n} \times [\mathbf{H}^+ - \mathbf{H}^-] \qquad \text{with} \quad r' \in \Gamma \tag{5}$$

$$J_s^m(r') = -\hat{n} \times [\mathbf{E}^+ - \mathbf{E}^-] \qquad \text{with} \quad r' \in \Gamma \tag{6}$$

The equivalent surfaces sources $J_s^e$ and $J_s^m$ are expressed in terms of the discontinuities of the total magnetic and electric fields, respectively, defined just outside ($\Gamma^+$) and inside ($\Gamma^-$) the contour surface $\Gamma$.

In Eq. (5) and (6) $\hat{n}$ indicates the outgoing normal unit vector to the surface $\Gamma$ at the point of definition of the equivalent source.

Moreover, suppression of the scattering outside the volume V is imposed, that allows deriving two compact conditions for the sources. It is worthwhile mentioning that as a function of the materials supporting the event, the presence of objects (dielectric, magnetic or conductive) in a given vacuum scenario can be equivalently described with different sources (volumetric and/or surface), since the equivalence by itself substitute the related effects of the constitutive parameters. Dielectric and magnetic materials, with permittivity $\varepsilon$ and permeability $\mu$ different from the vacuum parameters $\varepsilon_0$ and $\mu_0$, can be removed by considering the Volume Equivalence Principle (VEP) [31], whereas conductive objects that sustain surface discontinuities can be modeled in terms of the Surface Equivalence Principle (SEP) [32]. Even if the SEP is more general and it can include the VEP by considering a surface $\Gamma$ surrounding a certain volume V, the theoretical findings presented in this work show a single equivalent electromagnetic source in vacuum, obeying Maxwell's differential equations.

Applying the standard procedure to decouple the two Maxwell's equations, i.e. the *curl* operator to Eq. (1) and substituting the expression of the *curl* of the magnetic field, i.e., Eq. (2), and the same for Eq. (2) with substitution of Eq. (1), the expressions in Eqs. (7) and (8) are obtained

$$\nabla \times \nabla \times \mathbf{E} - k_0^2 \mathbf{E} = -i\omega\mu_0 \mathbf{J}_{eq} \tag{7}$$

$$\nabla \times \nabla \times \mathbf{H} - k_0^2 \mathbf{H} = \nabla \times \mathbf{J}_{eq} \tag{8}$$

with $k_0 = \omega\sqrt{\varepsilon_0\mu_0}$ denoting the (free space) wavenumber at the angular frequency $\omega$.

Due to the asymmetry in enforcing all the sources in the second member of Eq. (2), with their proper electric and magnetic effects according to the Eqs. (3-6), the source terms in Eqs. (7-8) are directly proportional to the equivalent sources and to the *curl* of them.

If the right-hand terms of Eqs. (7) and (8) vanish,

$$-i\omega\mu_0 \mathbf{J}_{eq}(\mathbf{r}') = 0 \tag{9}$$

$$\nabla \times \mathbf{J}_{eq}(\mathbf{r}') = 0 \tag{10}$$

i.e., the associated Helmholtz equation is considered that have solutions only if the rank of the system is reduced, no energy outside the volume V will be released, corresponding to a possible presence of trapped energy inside V (modes).

The two compact conditions in Eqs. (9) and (10) generalize the paradigm of strong and weak solutions for the cloaking in association with the condition for a source to be non-radiating [33]. While Eq. (10) is consistent with the result of Van Bladel [34], at a glance Eq. (9) is commonplace, at least for physical, i.e., realistic currents. However, considering that in the present investigation Eq. (9) refers to equivalent sources, it corresponds to a non-trivial condition, since it states that the nonexistence of (equivalent) sources determines nonappearance of scattering. In the following we are concentrate our attention on this aspect.

Considering Eqs. (5) and (6), the above observation reduces to:

$$\mathbf{E}^+ = \mathbf{E}^- \triangleq \mathbf{E} \tag{11}$$

$$\mathbf{H}^+ = \mathbf{H}^- \triangleq \mathbf{H} \tag{12}$$

In [23] it has been shown that considering the ratio(s) of the electric and magnetic field, one has an impedance $Z = E/H$ (or admittance $Y = H/E$) matching condition for each considered polarizations. Here, this approach is extended to show that there are other appearances, where Eq. (9) exhibits a nontrivial connotation in connection to nonradiating particles and in particular to anapole modes.

Firstly, let us assume the case when only volumetric sources are present, in which condition considering the explicit forms of the volumetric currents (in Eqs. (3) and (4)) Eq. (9) becomes

$$i\omega\varepsilon_0 \boldsymbol{P} - Y_0 \nabla \times \boldsymbol{M} = 0 \tag{13}$$

Equation (13) describes the required interaction between the polarization $\boldsymbol{P}$ and magnetization $\boldsymbol{M}$ vectors within a nonradiating particle. The result is consistent with the concept of null-field radiationless source [5] and it explains the Transformation Optics based approaches [18, 19] as well.

Going further with this assumption, by accepting that the magnetization corresponds to the *curl* of a toroidal dipole response in the volumetric system, i.e.

$$\boldsymbol{M} = \nabla \times \boldsymbol{T} \tag{14}$$

Eq. (13) can be revised as

$$\nabla \times \nabla \times \boldsymbol{T} - ik_0 \boldsymbol{P} = 0 \tag{15}$$

On the other hand, theorem IV of Deavaney-Wolf [8] states that if $\boldsymbol{F}$ is an arbitrary vector function that obeys to the condition

$$\nabla \times \nabla \times \boldsymbol{F} - k_0^2 \boldsymbol{F} = 0 \tag{16}$$

That means is nonradiating state.

Comparing Eqs. (15) and (16), a connection between $\boldsymbol{P}$ and $\boldsymbol{T}$ can be established achieving non-radiating configurations. In particular, assuming $\boldsymbol{F} \equiv \boldsymbol{T}$, one can see that for

$$\boldsymbol{P} = -ik_0 \boldsymbol{T} \tag{17}$$

Eq. (15) equals Eq. (16), i.e., the fields generated by $\boldsymbol{T}$ and $\boldsymbol{P}$ will cancel out, hence a non-radiating system has been obtained. Equation (17) is consistent with the result of [2], found for anapole mode establishment.

This essential fitting together between results coming from different research fields, e.g., optics and electromagnetics, described by Eq. (17) represents the main aspect of the present work, as concern the theoretical aspects. In the next section, we demonstrate how our approach can be adapted for realistic objects such as all-dielectric metamaterials, for example.

### III. VISUAL DEMONSTRATION OF NONRADIATING SOURCE

Based on equivalent sources approach described above, we introduce an example of nonradiating configuration due to symmetry breaking of a pseudo-toroidal structure. Let us assume for example a gedanken torus that possesses surface poloidal currents that occur due to interaction with incident electromagnetic wave. As can be seen from Fig. 2 (a), these currents create a loop of magnetic field within the torus and these modes combination generates a toroidal response $\boldsymbol{T}$ (see for example Eq. (14)).

Instead, if the central torus axis is moved from its initial position, it leads to a non-symmetric cross section within the torus (Fig. 2 (b)). Due to such symmetry breaking, local modification of the current path in the meridian planes are introduced. Consequently, in different planes unequable displacement currents $j_1$ and $j_2$ will be present. This geometrical asymmetry causes a spatial redistribution of the poloidal currents that create uncompensated electric response $P$ (not present in the initial configuration). Thus, the asymmetric torus possesses both electric and toroidal responses. These electric and toroidal moments both oscillate and at a given frequency, when Eq. (17) is satisfied, they give rise to a resonance. At this particular frequency, due to correct ratio between multipoles intensities, a dynamic anapole mode can be established; it is known as nonradiating state, characterized by a strong field concentration within the volume of the torus.

(a)                                                                                          (b)

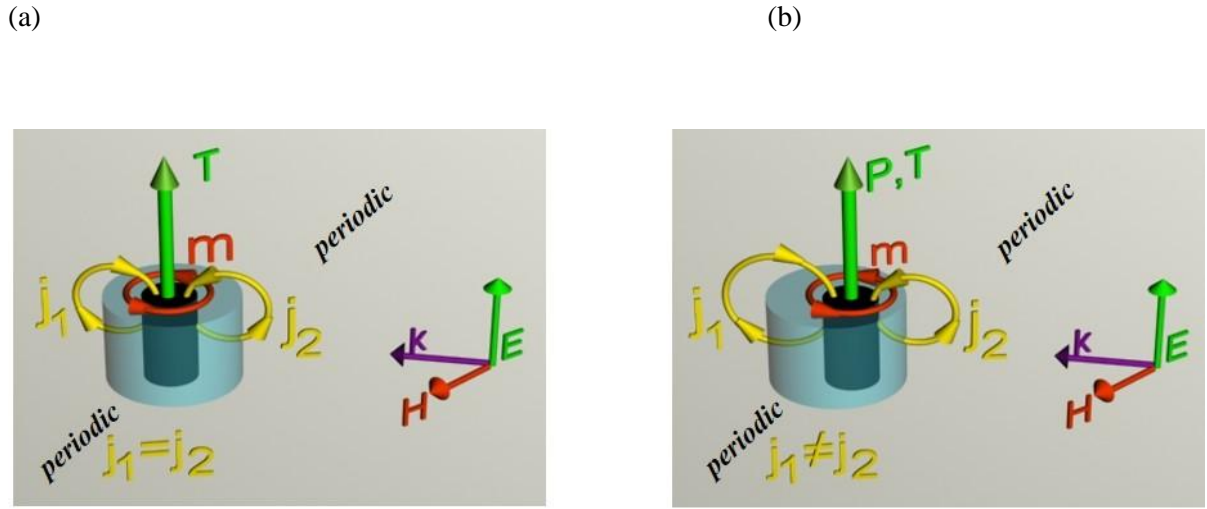

Fig. 2. Illustration of a metamolecule of subwavelength high-index dielectric cylinder with symmetric (a) and asymmetric (b) hole. Here $j$ denotes displacement currents, $m$ magnetic moment, $P$ and $T$ is electric and toroidal dipole moments, respectively.

Generally, the symmetric structure interacts with an incident electromagnetic wave and a single mode within the volume $V$ is generated, leading to a strong electromagnetic scattering. Here, within the modified configuration of electric and toroidal modes of the same order, which satisfies Eq. (17), with strongly suppressed electromagnetic scattering, i.e. non-radiating state is generated.

Here, we consider metamolecule (the unit cell of metamaterial), consisting of subwavelength high-index dielectric cylinder with centered hole, resembling a toroidal configuration (Fig. 2 (a)). An $E$ electric component of a linearly polarized incident wave is parallel to cylinder axis, creating so called magnetic Mie resonance that represents the electromagnetic response of dielectric particles in analogy with plasmonic resonance in metals [15, 36, 37]. According to Mie theory, the electromagnetic response of dielectric particle is underpinned by the displacement currents $j_1$=$j_2$, occurring within its volume by the incident wave. These displacement currents take the form of loops and generate aligned head-to-tail magnetic modes $m$ within the hole cylinder. Such electromagnetic configuration resembles the surface poloidal currents and magnetic moments within the torus that expectedly, create toroidal dipole $T$ moment oscillating along the symmetry axis [38-40].

On the other hand, if the hole is shifted from the center to any distance $d$, the produced asymmetry affects the charge-current distribution within a metamolecule (Fig. 2 (b)). Displacement currents $j$ still generate ring-like magnetic modes $m$ within the cylinder and their combination excites toroidal dipole $T$ moment. However, unequal redistribution of displacement currents $j_1$ and $j_2$ leads to uncompensated electric response $P$ that becomes resonant at certain frequency. Therefore, these electric $P$ and toroidal $T$ dipole moments oscillate in $P+ikT$ relation and precondition anapole mode establishment, satisfying equation (15) and, indeed, corresponds to nonradiating state [14].

The metamaterial consists of infinitely elongated $h=\infty$ dielectric cylinders placed in a vacuum. The radius of each cylinder is $R=5$ μm and of hole is $r=1$ μm with periodicity of metamolecules $l=12$ μm. In case of subwavelength dielectric particles we simulate properties of high-index dielectrics $\varepsilon=30$. We study electromagnetic response of metamaterial by commercial version of HFSS solver. The properly chosen parameters lead to couples magnetic Mie modes of metamolecules. We consider the cases of $d=0$ μm, 2 um, 3 μm, 3.5 μm and 4 μm hole displacements. To study the electromagnetic response of such metamaterial, a multipole decomposition of the displacement currents near resonant frequency range up to second order multipoles has been simulated, namely electric **P** and magnetic **M** dipole, toroidal **T** dipole, electric **Qe** and magnetic **Qm** quadrupole moments.

The centered hole position $d=0$ μm expectedly corresponds to dominating toroidal response of the system at f=1.215 THz (see Fig. 3 (a)). According to our approach, the asymmetric position of the hole with respect to the cylinder axis leads to redistribution of displacement currents $j_1$ and $j_2$ and, consequently, electric dipole response occurs. The gradual hole displacement for $d=2$ μm and 3 μm (Fig. 3 (b) and (c)) manifests in increasing of electric **P** dipole moment that becomes equal to toroidal **T** dipole moment for $d=3.5$ μm at f=1.215 THz (Fig 3 (d)). Eventually, electric dipole **P** moment intensity exceeds toroidal dipole **T** moment for $d=4$ μm (Fig 3 (e)).

However, the case of $d=3.5$ μm hole displacement is characterized by overlapping of the intensities of electric and toroidal dipole moments at f=1.215 THz corresponding to anapole mode establishment (Fig. 3 (d)) [11, 37].

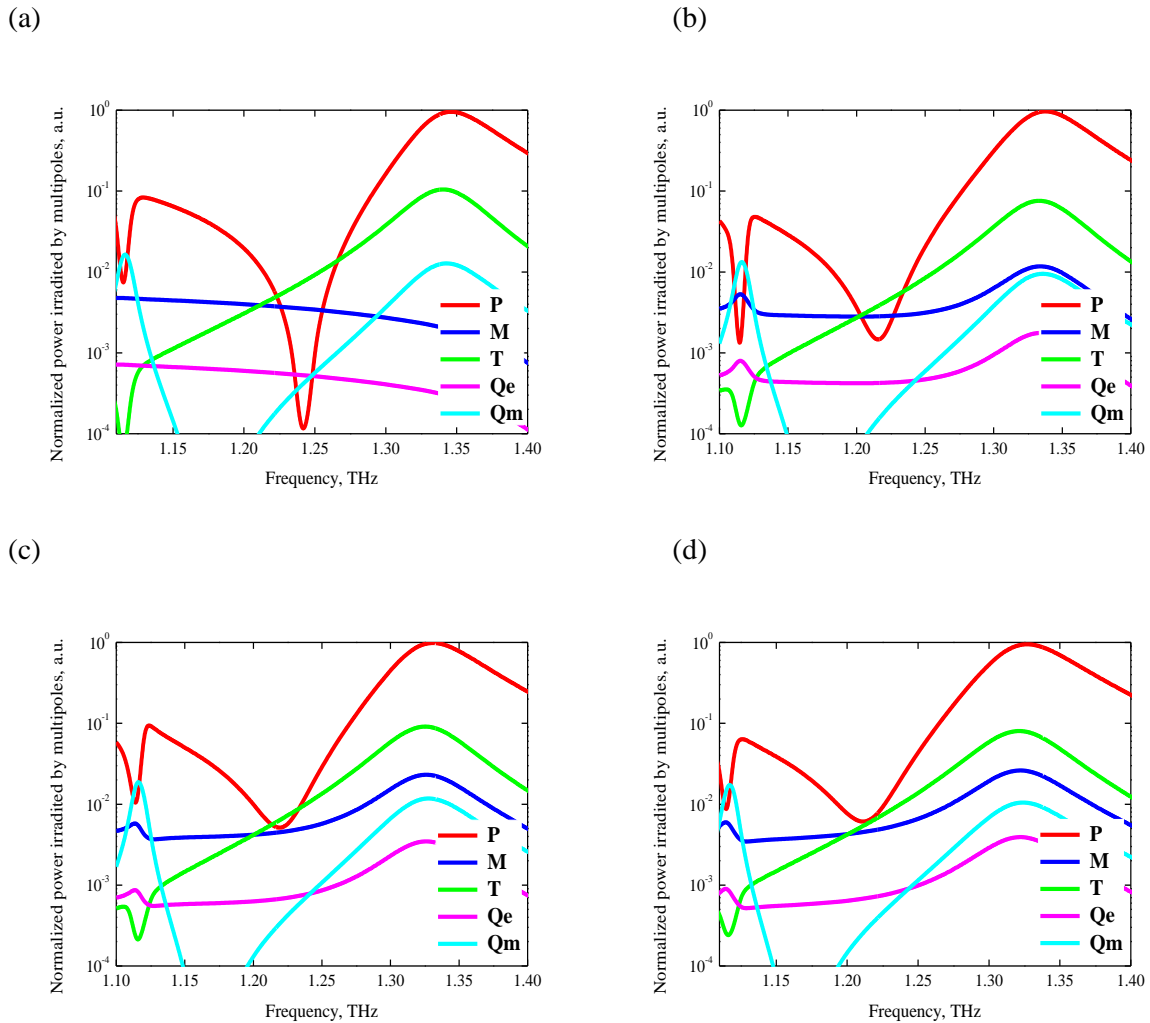

(e)

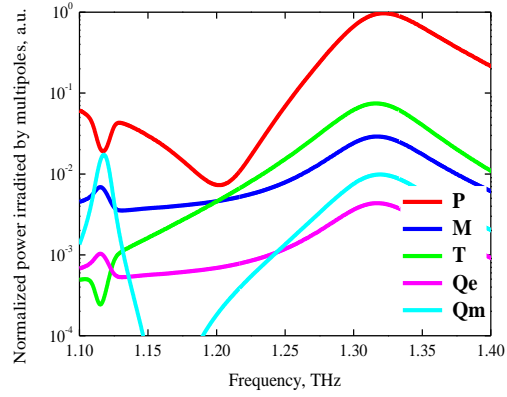

Fig. 3. Normalized power of displacement currents distribution of metamolecule in 1.1 THz-1.4 THz frequency range for d = 0 um (a), 2 um (b), 3 um, (c) 3.5 um (d) and 4 um (e) hole displacements.

Therefore, we analyze the electric and magnetic field distribution within metamolecule at this frequency. Fig. 4 shows the field maps of the absolute value of the electric and magnetic fields intensities. The electric field map features strong concentration in the center of the cylinder (Fig 4 (a)). Additionally, a strong magnetic field forms closed loop of magnetic field within metamolecule depicted at the same frequency at Fig. 4 (b). These fields correspond to anapole state establishment that possesses such strong field localization.

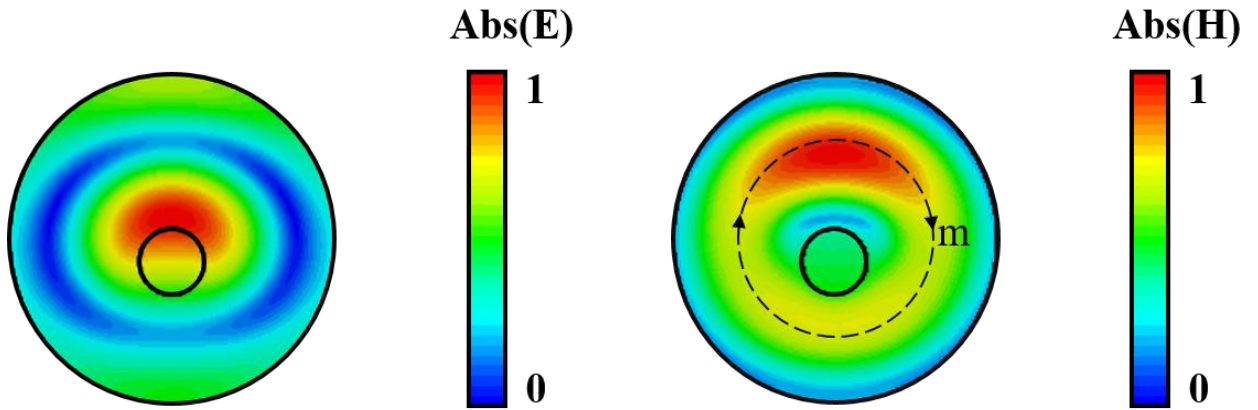

Fig. 4 Field maps of (a) y-component (along cylinders axis) of electric field and (b) absolute value of magnetic field intensities at f=1.215 THz.

In addition to strong field localization, anapole mode features suppressed scattering at resonance. In terms of multipole decomposition, it corresponds to strongly suppressed total radiation of electric dipole *P* and toroidal dipole *T* moments, i.e. *P*+*ikT*. Indeed, there is almost zero (0.000559) scattering losses at resonant frequency f=1.215 THz establishing anapole state (Fig. 5).

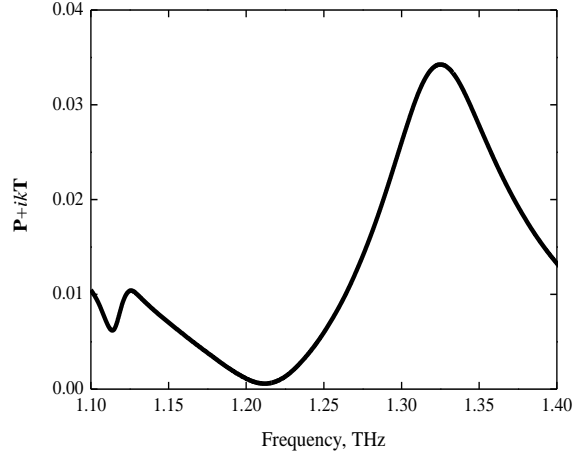

Fig. 5. Total radiation of electric dipole ***P*** and toroidal dipole ***T*** moments, i.e. ***P***+*ik****T*** for hole displacement *d*=3.5 μm at resonant frequency f=2.215 THz.

By means of this simple and clear example, we show that the symmetry breaking of toroidal structure establishes the anapole mode. Asymmetry configuration of the current distribution creates two oppositely directed current flows ***j***$_1$ and ***j***$_2$ that, in accordance with Eq. (17), are responsible for excitation of electric and toroidal dipole modes. At the same time, these currents satisfy Eq. (15), which corresponds to maintaining the nonradiating configuration of anapole state.

Apart from the electrodynamics in general, the formalism developed here will be of particular importance for the fields of anapole metamaterials, nanophotonics and invisibility. We formalize the general approach to the organization of nonradiating sources and scatterers. In particular, we declare that any nonradiating configuration are always result of two destructively oscillating modes coupling by a relation (16) of electric or magnetic origin. This idea can perfectly filled by electric and magnetic anapoles description [11, 14, 41]. Moreover, symmetry breaking of metaparticles is the second aspect leading to a non-radiating and invisible configuration which currently witness a surge of interest in the properties of transparent systems. Indeed, recently, a numbers of works have already confirmed the key role of anapole excitations in suppression of scattering properties of simple forms, such as all-dielectric nanodisks, clusters and SRR-hybrids. Correspondingly, one may desire to analyze of the electromagnetic response of structurally more complex scatterers for transparrency systems. Thus, our approach to asymmetric equivalent current sources could be useful for organizing asymmetrical (electromagnetically) metaparticles of anapole states.

## IV. CONCLUSION

In this work, unified theoretical findings on the common feature describing nonradiating sources are presented, in the framework of volumetric and surface equivalent sources. Consistently with the literature, theoretical results are found and, in addition, two compact conditions are reported for combining all the possible volumetric-volumetric, volumetric-surface and surface-surface configurations. Details and possible metamaterials or all-dielectric structures able to synthetize transparency devices and related anapole mode conditions in the volume *V*. Our approach will be useful for nonradiating states, anapole modes, transparent nanophotonics devices and metamaterials.

**ACKNOWLEDGMENTS**

The collaboration between Politecnico di Torino (Torino, Italy) and National University of Science and Technology - MISiS (Moscow, Russia) has been possible thanks to the project "Advanced Nonradiating Architectures Scattering Tenuously And Sustaining Invisible Anapoles" (ANASTASIA), funded by Compagnia di San Paolo in the framework of *Joint Projects for the Internationalization of Research*. This work was partly supported by the Ministry for Education and Science of the Russian Federation, in the framework of the Increase Competitiveness Program of the National University of Science and Technology MISiS under contract number and K2-2016-051, the Russian Foundation for Basic Research (Grant Agreements No. 16-32-50139 and No. 16-02-00789). The work on the multipoles decomposition investigation of the metamolecules was supported by Russian Science Foundation (project 17-19-01786).

In addition, G. L. and L. M. would like to acknowledge Luca Lussardi for stimulating mathematical discussions and the research group of Alexey Basharin for having provided ideal working conditions during their visiting period in Moscow.